# Author's Accepted Manuscript

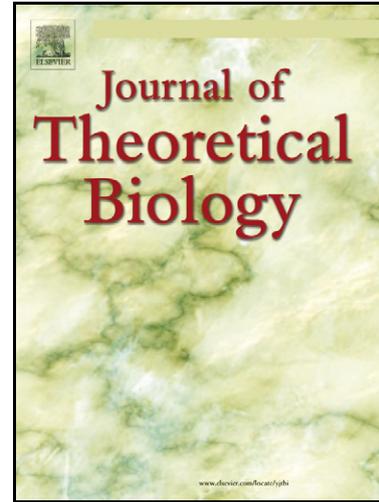

Stabilizing biological populations and metapopulations through Adaptive Limiter Control

Pratha Sah, Joseph Paul Salve, Sutirth Dey

Cite this article as: Pratha Sah, Joseph Paul Salve and Sutirth Dey, Stabilizing biological populations and metapopulations through Adaptive Limiter Control, *Journal of Theoretical Biology,* http://dx.doi.org/10.1016/j.jtbi.2012.12.014

**Title: Stabilizing biological populations and metapopulations through Adaptive Limiter Control**

**Authors: Pratha Sah[1]\*, Joseph Paul Salve\*\* and Sutirth Dey\*\*\***

**Affiliations:**

Population Biology Laboratory, Biology Division, Indian Institute of Science Education and Research-Pune, Pashan, Pune, Maharashtra, India, 411 021

\* ps875@georgetown.edu
\*\* josephsalve@gmail.com
\*\*\*Correspondence to Email: s.dey@iiserpune.ac.in

**1. Present address:** Department of Biology, Georgetown University, 37[th] and O Streets NW, Washington, DC, USA

**Name and address of the corresponding author:**
Sutirth Dey
Assistant Professor, Biology Division
Indian Institute of Science Education and Research
3[rd] floor, Central Tower, Sai Trinity Building
Garware Circle, Pashan
Pune - 411 021, Maharashtra, India
Tel: +91-20-25908054

1 **Abstract**

2 Despite great interest in techniques for stabilizing the dynamics of biological populations and

3 metapopulations, very few practicable methods have been developed or empirically tested. We

4 propose an easily implementable method, Adaptive Limiter Control (ALC), for reducing the

5 magnitude of fluctuation in population sizes and extinction frequencies and demonstrate its

6 efficacy in stabilizing laboratory populations and metapopulations of *Drosophila melanogaster*.

7 Metapopulation stability was attained through a combination of reduced size fluctuations and

8 synchrony at the subpopulation level. Simulations indicated that ALC was effective over a range

9 of maximal population growth rates, migration rates and population dynamics models. Since

10 simulations using broadly applicable, non-species-specific models of population dynamics were

11 able to capture most features of the experimental data, we expect our results to be applicable to a

12 wide range of species.

## 1. Introduction

Stabilizing the dynamics of unstable systems has been a major endeavor spanning different scientific disciplines. Unfortunately, most methods proposed in the literature require extensive *a priori* knowledge of the system and / or real-time access to the system parameters (Schöll and Schuster, 2008). This typically makes such methods unsuitable for controlling biological populations that are often characterized by poor knowledge of the underlying dynamics (however, see Suárez, 1999) and inaccessibility of the system parameters. This problem was partly alleviated with the advent of methods that needed no *a priori* knowledge of the system and perturbed the state variables rather than the system parameters (Corron et al., 2000; Güémez and Matías, 1993; Hilker and Westerhoff, 2007). For example, at least in single-humped one-dimensional maps, constant immigration of sufficient magnitude in every generation can convert chaotic dynamics into limit cycles (McCallum, 1992). Similar phenomena of simpler dynamics replacing more complex behaviour were also observed in models of more complex systems (e.g. (Astrom et al., 1996; McCann and Hastings, 1997). However, very few of these theoretical predictions have been empirically verified till date. In one experiment, the dynamics of *Tribolium* populations were stabilized by low magnitude perturbations (Desharnais et al., 2001). This method required the empirical characterization of the chaotic strange attractor of the dynamics, followed by computation of local Lyapunov exponents over the entire attractor: a somewhat daunting proposition for most application-oriented purposes. Another empirical study on a chemostat-based three-species bacteria-ciliate prey-predator system, implemented theoretically calculated rates of dilution to convert chaotic dynamics into limit cycles (Becks et al., 2005). Again, the calculations leading to the prediction of the dilution rates required fairly

1  detailed system-specific modeling (see Becks et al. 2005 and references therein) and were

2  implemented in a system that was spatially-unstructured.

4  One of the several complications with real populations is that they are very often spatially-

5  structured (metapopulations), which can lead to complex patterns and dynamics (Cain et al.,

6  1995; Maron and Harrison, 1997; Perfecto and Vandermeer, 2008; Turchin et al., 1998). Not

7  surprisingly therefore, the dynamics of metapopulations have received wide attention, in the

8  context of stabilization (e.g. Doebeli and Ruxton, 1997; Parekh et al., 1998). The rationale

9  behind such studies was that if the dynamics of a fraction of the subpopulations in a

10  metapopulation can be controlled in some way, then the stabilized subpopulations can alter the

11  dynamics of their neighbors and so on. Thus one could expect a cascading effect through the

12  metapopulation, ultimately leading to the stabilization of the global dynamics. However, the only

13  study using localized perturbations on real, biological metapopulations failed to find any effect

14  on global dynamics (Dey and Joshi, 2007). This was attributed to the effects of localized

15  extinctions in the subpopulations, which were shown to render a previously proposed method

16  (Parekh et al., 1998) ineffective in terms of stabilizing metapopulations. Thus, there are no

17  known methods that have been empirically demonstrated to stabilize the dynamics of biological

18  metapopulations.

19
20

21  One possible reason for this lack of empirical verification of proposed control methods

22  might be related to the multiplicity of notions related to population stability in ecology. Even 15

23  years back, a review on the subject had catalogued no less than 163 definitions and 70 concepts

24  pertaining to stability in the ecological literature (Grimm and Wissel, 1997). Most proposed control methods (Corron et al., 2000; Güémez and Matías, 1993; McCallum, 1992; Sinha and Parthasarathy, 1995; Solé et al., 1999) pertain to attainment of stability in the form of chaos being replaced by simpler dynamics (stable point or low periodicity limit cycles). While there have been a number of studies demonstrating chaos, or the lack there of, in empirical datasets (Becks and Arndt, 2008; Becks et al., 2005; Dennis et al., 1995; Hassell et al., 1976; Turchin and Taylor, 1992), many of the methods proposed for detecting chaos suffer from their own theoretical limitations (Becks et al., 2005; Turchin and Taylor, 1992). Moreover, the distinction between deterministic chaos and noisy limit cycles often does not lead to meaningful insights in terms of practical applications like resource management or reduction of the extinction probability of a population. Therefore, many experimental studies have concentrated on other attributes of stability that are relatively easier to determine, particularly in noisy systems. Two of the attributes of population stability often investigated in these contexts are the so called constancy (e.g. Mueller et al., 2000) and persistence (e.g. Ellner et al., 2001). A population is said to have greater constancy stability when it has a lower variation in size over time, while greater persistence stability simply refers to a lower probability of extinction within a given time frame (Grimm and Wissel, 1997). In this study, we empirically investigate both these attributes of population stability.

Here we propose a new method, which we call adaptive limiter control (ALC), for reducing the amplitude of fluctuation in population size over time. Our main motivation in proposing this method is to come up with a scheme that would be easy to implement, and at the same time, would be effective in terms of both constancy and persistence of spatially - unstructured and –structured populations. We first explore the method numerically and study its

1   long-term behaviour. We then use biologically realistic simulations (incorporating noise,

2   extinction and lattice effect) over a range of biologically meaningful parameter values to

3   demonstrate the efficacy of our method for populations with no migration (henceforth called

4   single populations) as well as spatially-structured populations experiencing migration among the

5   constituent subpopulations, henceforth called metapopulations (Hanski, 1999). We also report

6   two separate experiments using replicate single populations and metapopulations of *Drosophila*

7   *melanogaster* that validate our theoretical predictions. We further show that ALC reduces

8   extinction in both single populations and metapopulations, albeit by different mechanisms.

9   Finally, we compare ALC with other control methods in the literature, and point out why we

10  believe ALC to be likely applicable to a wide range of organisms.

1    **2. Adaptive Limiter Control (ALC) model**

2        Mathematically, ALC can be represented as:

3    $N_{t+1} = f(N_t)$                                   if $N_t \geq c \times N_{t-1}$,

4    $N_{t+1} = f(c \times N_{t-1})$                            if $N_t < c \times N_{t-1}$

6    where $N_t$ represents the population size at generation $t$, $f(N_t)$ is a function that predicts $N_{t+1}$ for a

7    given $N_t$, and $c$ is the ALC parameter. In other words, when the population size in the current

8    generation goes below a threshold, defined as a fraction $c$ of the population size in the previous

9    generation, individuals are added from outside to bring the number up to that threshold. No

10    perturbations are made if the population size is above that threshold. The biological

11    interpretation of this scheme is straightforward: the population size in the current generation (i.e.

12    $N_t$) is not allowed to go below a fraction $c$ of the previous population size ($N_{t-1}$). As the

13    magnitude of the control is a function of the population size in the previous generation, the

14    number of individuals added changes constantly. This adaptive nature of the algorithm makes it

15    independent of the range of the size of the populations to be controlled, thus enhancing its

16    applicability. ALC belongs to the so called "limiter control" family of algorithms (Corron et al.,

17    2000; Hilker and Westerhoff, 2006; Zhou, 2006), although to the best of our knowledge, this

18    particular scheme has not been proposed earlier in any context.

20        We began with an investigation of the effects of ALC on the steady-state behaviour of a

21    simple one-dimensional population dynamics model. As the calculation of the magnitude of

22    ALC involves population size over two generations, the dimensionality of the system is

23    increased, which makes precise analytical results difficult. Therefore, in this study, we limit

1    ourselves to numerical investigations of the effects of ALC. We used the widely-studied Ricker

2    map (Ricker, 1954) to represent the dynamics of the populations. This model is given as $N_{t+1} =$

3    $N_t \exp( r ( 1 - N_t / K ) ]$ where $N_t$, $r$ and $K$ denote the population size at time $t$, per-capita intrinsic

4    growth rate and the carrying capacity respectively. In the absence of any external perturbation,

5    this two-parameter model follows a period-doubling route to chaos with increase in the intrinsic

6    growth rate, $r$ (Fig 1A; May and Oster, 1976). In Fig 1 and Fig 2A, we studied the steady-state

7    behaviour by iterating the Ricker model in the absence of any noise for 1000 steps (larger

8    number of iterations did not lead to any qualitative changes in the graphs), and plotting the final

9    100 values. We also computed the fluctuation index (Dey and Joshi, 2006a) of the populations as

10   a measure of the corresponding constancy stability. The fluctuation index (FI) is a dimensionless

11   measure of the average one-step change in population numbers, scaled by the average population

12   size (see section 3.3.1 for details). As expected, when the population settles to a stable point

13   equilibrium, the FI is zero, but as the population enters the two-point limit-cycle zone, the FI

14   increases (Fig 1A). However, when the population becomes chaotic, the trajectory visits a large

15   number of points between the upper and lower bound, which can stabilize, increase or even

16   reduce the FI (Fig 1A). This demonstrates that there need not necessarily be a simple relationship

17   between the complexity of the dynamics and the corresponding constancy, and these two aspects

18   of stability are perhaps better addressed separately.

20        This point gets highlighted further when we consider the dynamics of the populations under

21   low levels of ALC ($c = 0.1$, Fig 1B) where the chaotic dynamics is replaced by simple limit

22   cycles, although the FI remains considerably high. In other words, at this level of ALC, whether

23   the population has been stabilized or not is a matter of interpretation in terms of the context of

1   the study. Increasing the magnitude of ALC (Fig 1C and 1D) restores the period doubling route

2   to chaos, although with a much reduced range of variation of population sizes. Comparing Fig

3   1A (no ALC) with 1B (low ALC), 1C (medium ALC) and 1D (high ALC), highlights that

4   although low ALC is able to ameliorate chaos effectively over a wide parameter range, medium

5   and high ALC are not. However, in terms of inducing constancy stability, medium and high

6   values of ALC are far more effective, even if they can not ameliorate chaos at these values.

7   These observations were substantiated by the bifurcation diagram at $r = 3.1$, and $c$ as the

8   bifurcation parameter (Fig 2A). Similar results were obtained using the logistic (May, 1974;

9   May, 1976) and the Hassell (Hassell et al., 1976) models, both of which have been used

10  extensively in the ecological literature for describing the dynamics of real populations. The

11  results obtained from these latter models are presented in the Supplementary Online Material

12  (Fig A4 and A5).

14      Small amounts of immigration stabilize the dynamics of most single-humped maps by

15  reducing the slope of the first return map at its point of intersection with the line of slope 1 (i.e.

16  $N_{t+1}=N_t$) (Stone and Hart, 1999). ALC also creates a floor for the values that the population size

17  can take and therefore does not allow the trajectory to visit certain parts of the attractor.

18  However, unlike in constant immigration, this floor is not a constant number, but keeps on

19  changing across generations, depending on the population sizes. As the value of $c$ increases, the

20  population size after perturbation tends towards the population size in the previous generation

21  (i.e. $c \times N_t \rightarrow N_{t-1}$). Since ALC is implemented only when there is a population decline (i.e. $N_{t-1}$

22  $>N_t$), it might seem that for high values of $c$, ALC can possibly lead to over-compensatory

23  dynamics, and hence increase the number and magnitude of population crashes. However,

1    looking at the bifurcation diagrams 1C and 1D, it is clear that on increasing the value of $c$, the

2    range of values of population sizes is reduced from both sides. This is also observed in Fig 2B,

3    where for $r = 3.1$, $K = 60$ and $c = 0.8$, we plot population sizes before applying ALC (pre-ALC),

4    after applying ALC (post-ALC) and the corresponding control ($c = 0$). Clearly, ALC reduces

5    both the number of crashes and the corresponding magnitude, even when the dynamics are

6    pushed into the over-compensatory zone. This is because, as long as $c < 1$, $N_{t-1} > N_t$ (post-ALC),

7    implying that the magnitude of the crash in next generation ($t+1$) is less than that in the previous

8    generation ($t$). This will automatically lead to reduction in the magnitude of fluctuations as well

9    as enhanced persistence.

11    To summarize, the magnitude of ALC to be used for a given population, can be

12    determined by the goal of the control process: lower values of ALC being employed for

13    ameliorating chaos, and medium to higher values for enhancing constancy. Most values of ALC

14    are expected to enhance persistence. Now we turn our attention towards real, biological systems

15    and investigate whether ALC can stabilize the dynamics of noisy, extinction-prone biological

16    populations and metapopulations.

1    **3. Materials and Methods**

2    **3.1 Biologically relevant simulations**

3        **3.1.1 The population dynamics model**

4        We continue to model the dynamics of single populations / subpopulations using the Ricker

5    (Ricker, 1954) equation. This is because first-principle derivations indicate that populations with

6    uniform random spatial distribution and scramble competition are expected to exhibit Ricker

7    dynamics (Brännström and Sumpter, 2005), and laboratory cultures of *Drosophila melanogaster*

8    (our model system) exhibit both properties. Moreover, prior empirical studies suggest that the

9    Ricker model is a good descriptor of the dynamics of single populations (Sheeba and Joshi,

10   1998) as well as metapopulations (Dey and Joshi, 2006a) of *Drosophila melanogaster*.

12   **3.1.2 Simulations incorporating biological / experimental realities**

13       *Noise and lattice effect:* Since real organisms always come in integer numbers (lattice

14   effect: Henson et al., 2001), we rounded off the model output at each iteration to the nearest

15   integer.  We also incorporated noise in the population growth rates in our simulations, by adding

16   a noise term $\varepsilon$ (-0.2<$\varepsilon$<0.2 uniform distribution) to $r$ in every generation. Thus, the final equation

17   used in our simulations can be represented as: $N_{t+1} = INT\ [N_t.\exp((r+\varepsilon).(1-N_t/K))]$; where INT

18   represents the integerization function. Based on estimates obtained by fitting the Ricker model to

19   some of the time series from the CTRL (i.e. $c = 0$) single populations, we fixed the value of $r$ as

20   3.1 in all the simulations. For the Ricker model, this value leads to chaotic dynamics (May and

21   Oster, 1976). The initial population size ($N_0$) and the carrying capacity ($K$) were fixed at 20 and

22   60 respectively.

1    ***Extinction and resets:*** The unmodified Ricker model never takes zero-values and therefore

2    is not suitable for modeling extinction-prone populations. On integerization of the population

3    size, the model does allow for extinction, but only at values of *r* which are substantially higher

4    than those observed in *Drosophila* cultures in the lab. Extinction is known to play a major role in

5    determining the dynamics of single populations and metapopulations (Dey and Joshi, 2006b;

6    Dey and Joshi, 2007), and hence was explicitly incorporated in our simulations. For this, we

7    stipulated a 50% chance of a population going extinct whenever $N_t < 4$ (Dey and Joshi, 2006a).

8    Following the experimental protocol, we reset the extinct CTRL single populations to a value of

9    8 in the next iteration. The CTRL metapopulations were reset only when both subpopulations

10    went extinct. In such cases, the value of each subpopulation was set to 8 in the next iteration.

12    ***Metapopulation simulations:*** All metapopulations consisted of two subpopulations with

13    symmetric rates of migration between them.

15    ***Imposition of ALC:*** In the metapopulation simulations, ALC was always imposed post

16    migration. Thus the influx due to ALC in a given generation would affect the population size of

17    the neighbor only by migration in the next generation. Following the experimental protocol, only

18    one subpopulation was subjected to ALC and the identity of the perturbed subpopulation was

19    maintained throughout the run.

21    ***Effects of noise and effort:***

22    All the simulations pertaining to the real populations (Figs 3-5) incorporated random noise,

23    drawn from a uniform distribution in the range ±0.2, in the intrinsic growth rate. We also

1    investigated whether changing these ranges can affect the performance of ALC in terms of

2    reducing the FI of single populations (Fig 6A). Furthermore, in a real-life scenario, it might not

3    be possible to apply the control value exactly in all generations. This might be due to

4    inaccuracies in the census, or just non-availability of the required number of organisms to be

5    added. To study the effect of such imprecision in the magnitude of ALC on the constancy

6    stability, we investigated scenarios where the value of $c$ was drawn randomly from uniform

7    distributions of different ranges, centered on a given value (X-axis of Fig 6B).

9    Whether a control method can be adapted in real life or not, depends to a large extent upon

10   the "effort" needed to apply the method. Following Hilker and Westerhoff (2005), we defined the

11   effort for a given value of ALC as the average number of individuals added to the population per

12   generation

13
$$Effort = [\sum_{t=t0}^{t=t0+T} abs(n_{ALC} - n_{ALC'})] / (\overline{n} \times T)$$

14   where $n_{ALC}$ is the population size after ALC imposition, $n_{ALC'}$ is population size without

15   ALC, T is the length of the time series and $\overline{n}$ is the average population size. In the above

16   equation, whenever no perturbation is imposed, $n_{ALC} = n_{ALC'}$, thus implying no effort for that

17   particular generation. Since the absolute values of the effort are expected to increase with $K$, we

18   have scaled it by the corresponding average population size. Evidently, the lower the values of

19   effort, the lesser number of individuals need to be added to the population on an average, which

20   will presumably be economically favorable.

22   ***Length of simulation runs and replicates:*** For all simulations except those on noise and

23   effort (Fig 6), we considered only the first 50 iterations for computing all statistics i.e. we

1 explicitly focused on the transients rather than the equilibrium behaviour. We consider this to be

2 more ecologically meaningful because while coupled map lattices can sometimes have very long

3 transient dynamics (super-transients) the environmental conditions of real populations are

4 unlikely to stay constant for that long (Hastings, 2004). All simulation runs were replicated 10

5 times and the corresponding means and standard errors reported.

7     For the simulations on noise (Fig 6A and B) and effort (Fig 6C), we rejected the first 900

8 iterations and computed the corresponding indices based on the subsequent 100 iterations. Thus,

9 we explicitly concentrate on the steady-state values for these quantities. The FI in figs 6A and 6B

10 were averaged over 100 replicate runs.

11     All simulations were performed using MATLAB ® R2010a (Mathworks Inc.).

13 **3.2 Experiments**

14     **3.2.1 Fly stocks used:** We used four large, out bred populations ($DB_{1-4}$) of the fruit fly *D.*

15 *melanogaster* derived from four long-standing laboratory populations called $JB_{1-4}$, whose

16 detailed maintenance regime has been documented elsewhere (Sheeba et al., 1998).

18     **3.2.2 Single population experiment:** We created six single-vial cultures from each DB

19 population, by placing precisely 20 eggs in 1.2 ml of banana-jaggery medium per 37 ml. vial.

20 Thus, each single population was represented by a vial culture in this experiment. Two single

21 populations derived from each DB population were randomly assigned to three treatments

22 namely CTRL ($c = 0$), and two levels of ALC: LALC ($c = 0.25$) and HALC ($c = 0.4$). The values

23 of $c$ were chosen based on the predictions derived from the simulations (see Fig 3A). We also

1. created twenty additional back-up populations, five each from $DB_{1-4}$, for generating the flies

2. needed for the ALC and CTRL resets (see below).

3.

4. The CTRL populations were maintained on a 21-day discrete generation cycle, following an

5. earlier protocol (Dey and Joshi, 2006a), for 15 generations (45 weeks). The larvae were raised on

6. 1.2 ml of standard banana-jaggery medium while the adults received excess live yeast paste in

7. addition to the food medium. Theoretical and empirical studies have shown that this combination

8. of larval and adult nutrition induces high-amplitude periodic oscillations in *D. melanogaster*

9. (Dey and Joshi, 2006a; Dey and Joshi, 2007; Mueller and Huynh, 1994; Sheeba and Joshi, 1998)

10. and, consequently, such populations often go extinct (i.e. no individuals in the population during

11. the census) (Dey and Joshi, 2006a). Whenever such extinctions happened, the CTRL populations

12. were reset by allowing 4 males and 4 females, randomly picked from the backup vials, to

13. oviposit for 24 hours. Strict ancestral correspondence was maintained while resetting the

14. populations.

15.

16. LALC and HALC single populations were maintained similar to the CTRL populations,

17. except the application of the appropriate ALC treatment immediately after census. To calculate

18. the magnitude of ALC to be applied in the $t^{th}$ generation, the population size of previous

19. generation (i.e. $N_{t-1}$) was multiplied by the limiter fraction, $c$ (= 0.25 or 0.4) and this product,

20. rounded off both ways to the nearest integer, formed the lower threshold for $N_t$. If $N_t$ was below

21. this threshold, the required number of impregnated female flies were added from the back-up

22. vials to bring make $N_t = c. N_{t-1}$. We used female flies for ALC since the dynamics of any sexual

23. species is chiefly determined by the number of females in the population (Dey and Joshi, 2006a;

1  Dey and Joshi, 2007). Since flies were being added each time the population size was low, the

2  ALC algorithm ensured that all extinct populations were automatically reset. However, unlike

3  the CTRL populations where the reset values are a constant (4 males + 4 females), in the ALC

4  populations, the resets happened with different numbers of individuals in each generation,

5  depending on the population size in the generation preceding the extinction.

8  **3.2.3 Metapopulation experiment:** For the metapopulation experiment, we initiated twelve

9  replicate metapopulations of *D. melanogaster*, from each of the ancestral DB$_{1-4}$. Each

10  metapopulation consisted of two subpopulations, each started by placing 20 eggs in 1.2 ml of

11  banana-jaggery medium. Thus, there were a total of forty-eight metapopulations comprised of 96

12  subpopulations, whose dynamics were monitored over 17 generations (51 weeks). The dynamics

13  of such two-patch metapopulations have been extensively investigated in the theoretical

14  literature (Gyllenberg et al., 1993; Hastings, 1993).

16  Sixteen of these metapopulations (four derived from each DB population) were randomly

17  assigned to the three treatments, namely CTRL ($c = 0$), and two levels of ALC: LALC ($c = 0.25$)

18  and HALC ($c = 0.4$). For each level of ALC, eight metapopulations (two from each DB)

19  experienced 10% migration between the subpopulations, while the remaining were subjected to a

20  migration rate of 30%. These rates of migration were chosen based on a previous study (Dey and

21  Joshi, 2006a) that had demonstrated that 10% migration induces metapopulation stability via

22  asynchrony, while 30% migration has the opposite effect. Each subpopulation was treated

23  exactly similar to the CTRL-single populations mentioned above till the point of census.

2 After census at each generation, migration was imposed by manually transferring the

3 required number of female flies between the subpopulations (Dey and Joshi, 2006a; Dey and

4 Joshi, 2007; Mueller and Joshi, 2000). To calculate the number of females to be migrated, the

5 census number was multiplied by the migration value (0.1 for 10% migration and 0.3 for 30%

6 migration) and then halved (assuming sex ratio to be 1:1) and rounded off both ways to the

7 nearest integer. ALC was imposed after census. Calculation of ALC followed the method

8 described in the single population section above, with the post-ALC population size of the

9 previous generation as the reference point. In case of the metapopulation with two

10 subpopulations, only one subpopulation was controlled throughout the duration of the

11 experiment.

13 A metapopulation was scored as extinct when there were no flies in either of the

14 subpopulations. Owing to the nature of the method, none of the ALC metapopulations needed to

15 be reset upon extinction. If any of the CTRL metapopulations went extinct, both subpopulations

16 were reset by adding 4 males and 4 females to 1.2 ml of media. A strict ancestral correspondence

17 was maintained during addition of adults for reset.

19 **3.3 Indices of stability and synchrony**

20 **3.3.1 Constancy:** Grimm and Wissel (Grimm and Wissel, 1997) proposed 'constancy' to

21 refer to the aspect of stability pertaining to amplitude of fluctuations in population size. We

22 measured constancy of metapopulations and single populations using the Fluctuation Index (FI)

23 which is the average one-step change in population size over generations (Dey and Joshi, 2006a):

2
$$FI = (1/T\bar{N}).\sum_{t=0}^{t-1} abs(N_{t+1} - N_t)$$

4    Here $\overline{N}$ is the mean population size over T generations and $N_{t+1}$ and $N_t$ are the population

5    size at generation t+1 and t respectively. Being a dimensionless quantity, the FI of any two

6    populations can be compared directly and higher values of FI indicate lower constancy stability

7    and vice-versa.

9    **3.3.2 Persistence:** We also investigated the persistence attribute of stability, which is the

10    opposite of the extinction propensity of a population. We quantified persistence in terms of

11    extinction frequency, measured as the number of times a metapopulation or population recorded

12    a population size of zero over the duration of the experiment. Thus, if a metapopulation in the

13    LALC treatment went extinct 4 times during the course of the experiment, the extinction

14    frequency was scored as 4 / 17 = 0.24 and so on.

16    **3.3.3 Synchrony:** Synchrony was calculated as the cross-correlation coefficient at lag zero

17    of the first differenced log transformed values of the two subpopulations sizes (Bjørnstad et al.,

18    1999).

20    **3.4 Statistical analyses**

21     For the single populations, the data were subjected to two-factor mixed-model ANOVA

22    with ALC (fixed factor, 3 levels: CTRL, LALC and HALC) crossed with ancestry (random

23    factor, 4 levels). The metapopulation data were analyzed in a three-way ANOVA framework

1  with ALC (3 levels: CTRL, LALC and HALC) crossed with migration (2 levels: low and high)

2  and ancestry ($DB_{1-4}$). Data analyses indicated no significant effects of ancestry ($DB_{1-4}$) or the

3  interaction of ancestry with any of the other factors. None of our statistical conclusions showed a

4  change when ancestry was included / excluded as a factor. We have chosen to retain the blocked

5  design here as that reduces the degrees of freedom of the denominator term in the F-ratio,

6  making our results more conservative. In all cases, Tukey's HSD was used for post-hoc tests of

7  significance for pair-wise differences among means. The extinction frequency data were arcsin-

8  square root transformed prior to analysis (Zar, 1999). All analyses were performed using

9  STATISTICA ® v 9.1 (Statsoft Inc.).

1    **4. Results**

2    **4.1 Effect of ALC on single populations**

3        The Ricker-based simulations predicted a general reduction in population FI (and hence

4    enhanced constancy) on increasing the strength of ALC for a single population (Fig 3A). The 15-

5    generation experiment on the effects of ALC on constancy of single populations corroborated the

6    simulations, and we found a significant effect of the control magnitude $c$ ($F_{2,6}$ = 25.231,

7    $P<0.001$; LALC<CTRL, $P<0.007$; HALC<LALC, $P<0.004$), with the LALC ($c$ = 0.25) and the

8    HALC ($c$ = 0.4) populations experiencing a mean reduction of 19.03% and 39.89% in the FI (Fig

9    3B). This reduction can also be visualized in the time series of the control and treatment

10   populations (see Appendix Fig A1). ALC also reduced the extinction frequency of single

11   populations, though the effect was not statistically significant at α = 0.05 ($F_{2,6}$ = 2.6426, $P<0.15$;

12   Fig 3C).

14   **4.2 Effect of ALC on metapopulations**

15       **4.2.1 Simulations:** We next investigated the effect of ALC in metapopulations with two

16   subpopulations, where the dynamics of each subpopulation was governed by a stochastic Ricker

17   equation. We tested the stability of the system both at the global (metapopulation FI) and local

18   (subpopulation FI) level. The simulations indicated that when migration rate is high (30%), there

19   is a monotonic reduction in metapopulation FI with increase in $c$, although the rate of decrease

20   reduces after $c \sim 0.2$ (Fig 4A). On the other hand, for a low rate of migration (i.e. 10%), ALC

21   initially reduces the metapopulation FI, which stays more or less the same up to $c$ = 0.2. After

22   this point, at low migration rates, the metapopulation FI increases slowly, although it never

23   makes the population more destabilized than the controls. This somewhat anomalous behaviour at 10% migration level can be explained in terms of the synchrony between the subpopulations. Low and high migration are known to induce negative and positive correlation respectively between the subpopulations (Dey and Joshi, 2006a). At larger values of $c$, ALC reduces the magnitude of correlation between the subpopulations (Fig 5A). This is expected since medium to large values of $c$ do not ameliorate chaos, whereas smaller values of $c$ lead to limit cycles (Fig 2A). In terms of synchrony, chaotic behaviour is predicted to reduce the magnitude of the correlation coefficient between the subpopulations, while limit cycles is likely to enhance it. Thus, for both migration rates, the magnitude of the correlation coefficient tends towards zero (i.e. no correlation) with increase in $c$. However, this is expected to have contrasting effects on the metapopulation FI. This is because while reducing the positive synchrony among subpopulations reduces metapopulation FI, a decrease in negative synchrony among the subpopulations has the opposite effect (Dey and Joshi, 2006a; Hastings, 1993). This explains why, with increasing $c$, the metapopulation FI reduces for high migration rates, but tends to increase for low migration rates. However, crucially, even under low migration rate, the metapopulation FI never goes beyond the control, suggesting that there is no net destabilization due to ALC. At the subpopulation (or local) level, ALC populations showed reduced FI compared to the controls across both low and high migration rates (Fig 4C). The reduction in subpopulation FI was constant and independent of the ALC magnitude $c$, indicating that the differential impact of ALC on metapopulation constancy is expected to be modulated through its ability to alter the synchrony between the subpopulations.

**4.2.2 Experiment: Metapopulation constancy:** In the experiment, there was a significant main effect of ALC on the metapopulation FI ($F_{2,6} = 6.0265$, $P<0.037$; LALC<CTRL,

1  $P<0.01$; HALC<CTRL, $P<0.002$) with the LALC and the HALC experiencing a 15.52 % and

2  19.54 % reduction compared to the controls, but no significant difference ($P<0.69$) between the

3  two ALC treatments (Fig 4B). This shows that while ALC reduces metapopulation FI, increasing

4  the magnitude of $c$ does not lead to greater stability. High rate of migration is known to be a

5  destabilizing factor (Dey and Joshi, 2006a), and as expected, we found an almost significant ($F_{1,3}$

6  = 7.5886, $P<0.07$) effect of migration on metapopulation FI (Fig 4B). However, there was no

7  significant interaction between the migration rate and ALC ($F_{2,6}$=0.7582, $P<0.51$), suggesting

8  that the application of this method may not require a priori estimations of the migration rate.

10  **4.2.3 Experiment: Subpopulation constancy:** Locally, there was a significant main

11  effect of ALC ($F_{2,6}$ = 21.026, $P<0.002$; LALC<CTRL, $P<0.0002$; HALC<CTRL $P<0.0001$) and

12  migration ($F_{1,3}$= 11.274, $P<0.0438$) on subpopulation FI (Fig 4D), but no ALC × migration

13  interaction ($F_{2,6}$=0.23, $P<0.80$). Thus, as predicted (Fig 4C), ALC stabilizes subpopulations, but

14  does not lead to greater stability ($P<0.29$) on increasing the value of $c$ (Fig 4D).

16  **4.2.4 Experiment: Synchrony:** There was a significant main effect ($F_{1,3}$=17.59, $P<0.02$)

17  of migration, but not ALC ($F_{2,6}$=2.17, $P<0.19$), on the synchrony between the subpopulations

18  (Fig 5B). Although there was a substantially greater reduction in synchrony of ALC populations

19  under high migration compared to low migration (Fig 5B), the ALC × migration interaction was

20  found to be non-significant ($F_{2,6}$=0.84, $P<0.47$). Reduced synchrony among subpopulations is

21  known to decrease metapopulation fluctuations (Dey and Joshi, 2006a; Gyllenberg et al., 1993;

22  Hastings, 1993). Thus, our findings on synchrony are consistent with the results on

1     metapopulation stability (Fig 4B), where there was a greater reduction in metapopulation FI

2     under high migration compared to the low migration.

4     **4.2.5 Experiment: Extinctions:** Reduction in synchrony among the subpopulations has

5     been predicted to enhance metapopulation persistence (Ben-Zion et al., 2011; Heino et al., 1997).

6     We found an almost significant main effect of ALC on the extinction probability of the

7     metapopulations ($F_{2,6}$=4.93, $P$<0.054; Fig 5C), but neither a significant effect of migration

8     ($F_{1,3}$=5.2374, $P$<0.11) nor a migration × ALC interaction ($F_{2,6}$=0.67, $P$<0.94). Thus, ALC was

9     seen to enhance both constancy and persistence of single and metapopulations.

11     **4.3 Effects of noise and effort**

12     Although we have demonstrated the efficacy of ALC in laboratory populations and

13     metapopulations, the effectiveness of the method under natural settings will depend upon, *inter*

14     *alia*, its robustness to noise. ALC was found to be robust to moderate to high levels of noise in

15     the intrinsic growth rate (*cf* Fig 6A and Fig 2A). More importantly, even when there were noise

16     in *c*, ALC was able to increase the constancy stability of the populations (Fig 6B). Together,

17     these indicate that ALC is likely to stabilize populations even when there is considerable noise in

18     the system (which would ultimately reflect as noise in the growth rate), or the control is not

19     imposed with high degree of accuracy. The latter criterion is particularly important from an

20     applicability point of view, since ALC depends upon introduction of individuals from external

21     sources or populations, which, in practice, can some times be unreliable!

22     The effort needed to implement ALC was maximal at intermediate values of *c*, a phenomenon

23     that deserves future theoretical exploration. However, more important from the perspective of

1   this study, we find that even the maximum effort was less than 30% of the average population

2   size (Fig 6C), indicating that the number of organisms needed to stabilize the populations would

3   not be prohibitively large. Unfortunately, previous studies that have reported the efforts (Dattani

4   et al., 2011; Hilker and Westerhoff, 2005), use slightly different versions of the Ricker model,

5   and corresponding parameter values, rendering direct comparisons across studies difficult.

7   **5. Discussion**

8   **5.1 Effect of ALC on single populations**

9   In order to be applicable to a real population, the method needed to be robust to

10  biologically realistic scenarios. Since ALC was found to be effective in simulations

11  incorporating extinctions, noise and integerization of the state variable (Fig 3A), it was a likely

12  candidate for controlling real populations that typically exhibit these features. As predicted by

13  the simulations, we found a significant effect of ALC, on constancy (FI) in our single

14  populations experiment (Fig 3B).  Interestingly, we also found a slight increase in the average

15  population size of the ALC treatments (HALC: 32.96±1.25; LALC: 32.10±1.69; CTRL:

16  29.91±1.25), an effect that has been previously predicted for the so called "limiter from below"

17  method (Hilker and Westerhoff, 2005). This large reduction in FI in LALC and HALC, coupled

18  with greater average population size in LALC and HALC than the CTRL, suggested that ALC

19  might also reduce the probability of extinction. This is because populations experiencing lower

20  FI and greater average size are expected to hit lower population values less frequently over time,

21  and hence are expected to be less prone to extinction due to demographic stochasticity (Dey et

22  al., 2008). This prediction was verified qualitatively when we observed that ALC reduced the

23  extinction frequency of single populations (Fig 3C). Taken together, these observations

1  suggested that ALC might be a candidate method for stabilizing real metapopulations via

2  localized perturbations, both in terms of constancy and persistence.

4  **5.2 Effect of ALC on metapopulations**

5  **5.2.1 Constancy:** In principle, the dynamics of a metapopulation can be stabilized by

6  applying an appropriate perturbation to each subpopulation (Parekh and Sinha, 2003). However,

7  such a scheme is difficult to implement, and therefore most studies focus on stabilizing

8  metapopulations through localized perturbations to a subset of the subpopulations (Doebeli and

9  Ruxton, 1997; Parekh and Sinha, 2003; Solé et al., 1999). Moreover, migration is also known to

10 play a significant role in determining metapopulation stability (Ben-Zion et al., 2011; Dey and

11 Joshi, 2006a; Hastings, 1993) and perturbation and migration can interact in complex ways to

12 determine the dynamics of spatially-explicit systems (Singh et al., 2011; Steiner et al., 2011).

13 Therefore, in this study, we explicitly looked at the interaction of migration rates and ALC on the

14 dynamics of metapopulations. The significant effect of migration on FI (Fig 4B) was expected as

15 it is known that low (10%) migration reduces metapopulation FI (Dey and Joshi, 2006a).

16 However, the crucial observation here is the lack of interaction between migration and ALC,

17 which indicates that precise information about the migration rate may not be necessary before

18 applying ALC.

20   Although ALC reduced subpopulation FI at both levels of migration (Fig 4D), there was

21 a larger net reduction in metapopulation FI under the high migration rate (Fig 4B). This is

22 intuitive, since prior theoretical (Gyllenberg et al., 1993; Hastings, 1993) and empirical (Dey and

23 Joshi, 2006a) studies indicate that lower rates of migration stabilize metapopulations to a great

1   extent. Therefore, the stability induced by ALC is likely to be more prominent only when the

2   metapopulations are relatively unstable to begin with, which in this case was due to high rates of

3   migration. Importantly, since ALC did not destabilize the stable metapopulations (i.e. the 10%

4   migration treatment), it follows that the method can be applied even in the absence of precise

5   estimates of the migration rates without risk of destabilization.

7       Interestingly, the simulations also suggest that the degree to which

8   metapopulation/subpopulation FI decreases due to ALC is not grossly affected by the magnitude

9   of $c$ (Fig 4A, 4C). This indicates that although moderate levels of $c$ are expected to be effective

10  in reducing metapopulation/subpopulation fluctuations, increasing the magnitude of the control

11  may not affect constancy any further, an observation corroborated in the metapopulation

12  experiment (Fig 4B, 4D). The non-significant difference between metapopulation FI of LALC

13  and HALC populations suggest that while LALC was enough to stabilize the metapopulations,

14  no significant gains were obtained by increasing the magnitude of the control.

16  **5.2.2. Synchrony:** Although there was a substantial reduction in the synchrony of ALC

17  populations under high rates of migration (Fig 5B), the ANOVA showed no significant effect of

18  ALC on synchrony. This might be due to the relatively large variation around the mean of the

19  CTRL populations undergoing low migration, in turn attributable most probably to experimental

20  noise. In fact, analyzing the synchrony data separately for the two migration rates show a

21  significant effect of ALC under high ($F_{2,6}$=7.0615, $P$<0.027) but not low migration ($F_{2,6}$ =0.126,

22  $P$<0.88), suggesting that ALC might cause a greater reduction in synchrony under high

1   migration. This is once again intuitive since low migration alone can significantly reduce the

2   subpopulation synchrony (Dey and Joshi, 2006a; Gyllenberg et al., 1993; Hastings, 1993).

4   **5.2.3 Persistence:** Reduction in subpopulation synchrony is also expected to enhance

5   metapopulation persistence (Ben-Zion et al., 2011; Heino et al., 1997). This is because

6   subpopulations that are out of sync with each other have a greater chance of a locally extinct

7   population to be recolonized by immigrants from a neighboring population. Based on the

8   empirical observations on synchrony alone, one would expect a slightly larger reduction in the

9   extinction frequency of the ALC metapopulations (relative to the CTRL) under the high, but not

10  the low migration regime. However, we found that ALC reduced the extinction frequency under

11  both migration rates to a similar degree (Fig 5B). This somewhat anomalous reduction of

12  extinction frequency of the ALC treatments under low migration was possibly due to the

13  decrease in subpopulation FI (Fig 4D), which is also known to reduce the extinction probability

14  of metapopulations (Griffen and Drake, 2009). Expectedly, across all three treatments, the lower

15  extinction frequency in the metapopulations undergoing low rates of migration was accompanied

16  by corresponding lower levels of synchrony, compared to the high migration treatments (Ben-

17  Zion et al., 2010; Ben-Zion et al., 2011).

19  **5.3 Comparison with other methods:**

20  ALC has several advantages compared to some of the other methods investigated in the

21  control literature. Unlike the OGY method (Ott et al., 1990) and its variants (reviewed in

22  Andrievskii and Fradkov, 2004; Kapitaniak, 1996; Schöll and Schuster, 2008), computation of

23  the perturbation value does not require precise knowledge of either the dynamics and / or real-

1    time estimation of the parameters thereof. This makes ALC more suitable for application for

2    biological populations, where the underlying models are typically not known reliably and the

3    parameters (like intrinsic growth rate or carrying capacity) are often derived *a posteriori* through

4    model-fitting, and thus are not accessible in real time. ALC belongs to the class of methods in

5    which the state variables are directly perturbed to attain the desired level of control. Such

6    methods typically advocate the setting of a threshold below / above which the population is not

7    allowed to venture (Hilker and Westerhoff, 2005; Hilker and Westerhoff, 2006; McCallum,

8    1992; Sinha and Parthasarathy, 1995; Solé et al., 1999). These methods work by not allowing the

9    population numbers to attain extreme values that can lead to large increase or decrease (boom or

10   bust) in the next generation (Stone, 1993; Stone and Hart, 1999). Another variant of these

11   methods require the identification of the pre-images of the crashes in the time series, and perturb

12   the populations each time they enter into such "alert zones" (Desharnais et al., 2001; Hilker and

13   Westerhoff, 2007). However, such methods are invariably saddled with the problem of a *priori*

14   decision in terms of the thresholds or alert zones below / beyond / at which the perturbations

15   need to be made. This is a major issue with natural populations wherein the carrying capacity of

16   the environment and the intrinsic growth rates of the same species are liable to vary between

17   populations, which implies that the value of the control threshold has to be determined on a case-

18   by-case basis, through prior knowledge of the dynamics of the given population. Moreover,

19   natural populations might exhibit increasing / decreasing trends in size (Turchin, 2003) due to

20   extrinsic factors, which would make determination of the threshold even more problematic. This

21   problem is partially alleviated by another class of methods  in which the perturbations are not

22   hard numbers, but proportionate to some quantity, usually the present population size (Doebeli

23   and Ruxton, 1997; Güémez and Matías, 1993; Solé et al., 1999) or the difference between the present population size and some pre-determined threshold (Dattani et al., 2011). This makes the magnitude of the perturbation "adaptive" to the present population size, and therefore likely to be more useful for real populations. ALC belongs to the class of proportionate feedback methods, and like other members of the class, requires the *a priori* estimation of the proportion to be perturbed. However, as we have already shown, the performance of ALC does not change much over a relatively large range of values of $c$, which reduces the need for precise guesses about the values of $c$ to be used for real populations.

## 6. Concluding Remarks

Since we demonstrate the efficacy of ALC using *Drosophila* populations, it is natural to ask whether this method will be applicable to other species as well. As simulations using the non-species-specific Ricker model were able to capture most of the qualitative features of the dynamics, we believe that our results do not depend upon any idiosyncratic feature of *Drosophila* laboratory ecology. Moreover, the Ricker map has been applied to model the dynamics of organisms as diverse as bacteria (Ponciano et al., 2005), fungi (Ives et al., 2004), ciliates (Fryxell et al., 2005) crustaceans (Drake and Griffen, 2009), fruit flies (Cheke and Holt, 1993; Sheeba and Joshi, 1998), fishes (Ricker, 1954) etc. This is probably because first-principle derivations indicate that populations with uniform random spatial distribution and scramble competition are expected to exhibit Ricker dynamics (Brännström and Sumpter, 2005), and natural / laboratory populations of several organisms typically satisfy these two conditions. Thus, predictions arising out of this generic and widely-applicable model are likely to be relevant across several taxa. Our results on the Ricker map were found to be valid for two other commonly used models of population dynamics, namely the Logistic (May, 1974) and Hassell

1   (Hassell et al., 1976) equations (See Supplementary Online Material Fig A4 and A5). Thus, ALC

2   is likely to be effective across a wide range of systems.

4       Stabilizing the dynamics of populations in terms of constancy and / or persistence is a

5   major concern for conservation biologists and ALC seems to be effective on both counts. This is

6   notable from a possible application point of view, as constancy and persistence do not

7   necessarily correlate (Dey et al., 2008). Furthermore, ALC ensures a constant genetic influx,

8   which is an important component of maintaining genetic variation in a population (Biebach and

9   Keller, 2012), although see (Heath et al., 2003). However, before applying ALC to other

10  systems, one should be aware of some of the caveats of the present study. Our experiments and

11  simulations pertain to organisms with high population growth rates and exhibiting Ricker /

12  Logistic / Hassell type of discrete dynamics, whereas many organisms of concern to conservation

13  biologists (e.g. mammals and birds) often have much lower growth rates, and qualitatively

14  different life-history and dynamics from the kind that we have considered here (Fronhofer et al.,

15  2012). Moreover, metapopulation dynamics are known to depend upon migration schemes (Earn

16  et al., 2000) and the precised nature of density-dependence (Ims and Andreassen, 2005), two

17  factors that we have not considered in our study. Thus, any extrapolation of ALC to a different

18  biological population should be tempered with caution and relevant system-specific information.

1  **ACKNOWLEDGEMENTS**

3    We thank Shraddha Karve and Arun Neru for assistance in the laboratory. Frank Hilker

4  and an anonymous reviewer provided several constructive suggestions pertaining to the

5  theoretical aspects of the study. This work was supported by an extra-mural grant from Council

6  for Scientific and Industrial Research, Government of India. We also thank IISER-Pune for

7  financial support through in-house funding.

1    **Figure Legends**

3    **Figure 1. Effect of ALC on the dynamics of an uncoupled Ricker map.** Simulations with $N_0$

4    =20 and $K$ = 60. **(A)** An unperturbed ($c$ = 0) isolated single population shows classic period-

5    doubling route to chaos. FI increases rapidly from zero as the Ricker map undergoes period

6    doubling bifurcations to chaos. FI shows an irregular but a gradual increase at high values of $r$

7    (>3.0) **(B)** At $c$ = 0.1 the bifurcation map is largely reduced to a two point cycle, thus converting

8    the complex dynamics of the system to simple two-point limit cycle. **(C)** $c$ = 0.4 and **(D)** $c$ =

9    0.75 further reduce the amplitude of population size fluctuations in the bifurcation map as well as

10   reduce the maximum FI reached by the population. Note that the decrease in the amplitude and

11   reduction in FI is proportional to the increasing value of $c$.

13   **Figure.2. Effect of magnitude of ALC value in an uncoupled Ricker map (A)** Simulations

14   over 1000 iterations, taking Ricker growth rate parameter, $r$ = 3.1 and $K$ = 60. Increasing the

15   value of $c$ decreases the amplitude of population size fluctuation for a single population. FI

16   decreases monotonically with $c$ except for very small values ($c$ = 0.05) where there is a slight

17   increase. Although ALC does not lead to limit cycles beyond $c$ = 0.3, it does cause a reduction in

18   FI, thus enhancing constancy. This highlights that constancy and simpler dynamics do not

19   necessarily correlate, and the values of ALC to be used should depend on the kind of stability

20   desired in the system. **(B)** Simulated time series showing pre-ALC and post-ALC values. Here

21   we visualize the effect of ALC on the same time series, before and after the application of the

22   perturbation. Clearly, ALC reduces the magnitude of fluctuation even in the pre-ALC time

23   series. The completely unperturbed series is represented by $c$ = 0.

**Figure 3. Effects of ALC on a single populations**. Simulations averaged over 50 iterations and 10 replicates, taking Ricker growth rate parameter, $r$ = 3.1. Each experimental data-point is the mean of 8 replicate single populations. Error bars denote ± SEM. **(A)** Simulations showed that population FI decreases with increase in the magnitude of the ALC parameter, $c$. **(B)** Experiments showed that mean population FI decreases with increasing values of $c$. Note that the scale on Y-axis is different from 3A. (C) ALC reduces extinction frequency in experimental populations. Comparing the simulations and the experimental results in A and B indicate a good correspondence between the two.

**Figure 4: Effects of ALC on metapopulation and subpopulation constancy. (A)** Simulations: Metapopulation FI decreases with increasing $c$ at high migration. At low migration, metapopulation FI is lowest at intermediate values of $c$. **(B)** Experiment: Although metapopulation FI decreases significantly at high migration, there was no significant effect of ALC under low migration. **(C)** Simulations: Compared to an unperturbed ($c$=0) system, ALC reduces the subpopulation FI at both migration rates. **(D)** Experiments: Under low migration, there was a significant reduction in subpopulation FI, which might have contributed to the enhanced persistence. Along with reduced synchrony (Fig 5B), the significant reduction in subpopulation FI under high migration might also have contributed to the greater persistence of these metapopulations. Error bars denote ± SEM.

**Figure 5: Effects of ALC on metapopulation synchrony and persistence. (A)** Simulations: Increasing ALC reduces the magnitude of the cross-correlation coefficient between the subpopulations under both rates of migration. **(B)** In the experiments, ALC significantly reduces

1   the synchrony among the subpopulations under high level of migration. However, under low

2   migration, there was no significant effect of ALC on synchrony (see section 5.2.2 for

3   discussion). This leads to the prediction that there would be difference in terms of persistence

4   under the high, but not the low migration treatments. **(C)** ALC was found to be effective in

5   reducing the extinction frequency under both high and low migration (see section 5.2.3 for a

6   possible explanation).

8   **Figure 6: Simulations on effects of noise on ALC and the effort. (A)** Adding various levels of

9   noise to the value of intrinsic growth rate parameter $r$ does not affect the performance of ALC in

10  terms of enhancing constancy stability **(B)** Noise in ALC magnitude $c$ does not affect its ability

11  to reduce the FI of the system. These simulations indicate that ALC is a robust method for

12  attaining population stability. Each point in these two figures represent FI computed over 100

13  iterations (after rejecting 900 transients) and are averaged over 100 replicates, **(C)** The

14  maximum scaled average effort (see section 3.1.2 for explanation) of implementing ALC does

15  not exceed the value of 0.3. This implies that not too many individuals need to be added to

16  implement ALC.

18  **Highlights**

19  ♦  Novel method for controlling the dynamics of populations/metapopulations.
20
21  ♦  First control method empirically shown to work for a biological metapopulation.
22
23  ♦  The method reduces both population fluctuations as well as extinction probability.
24
25  ♦  Biologically realistic simulations indicate the results to be widely applicable.
26  ♦  We provide empirical validation for various extant theoretical studies in this area.

4. Figure

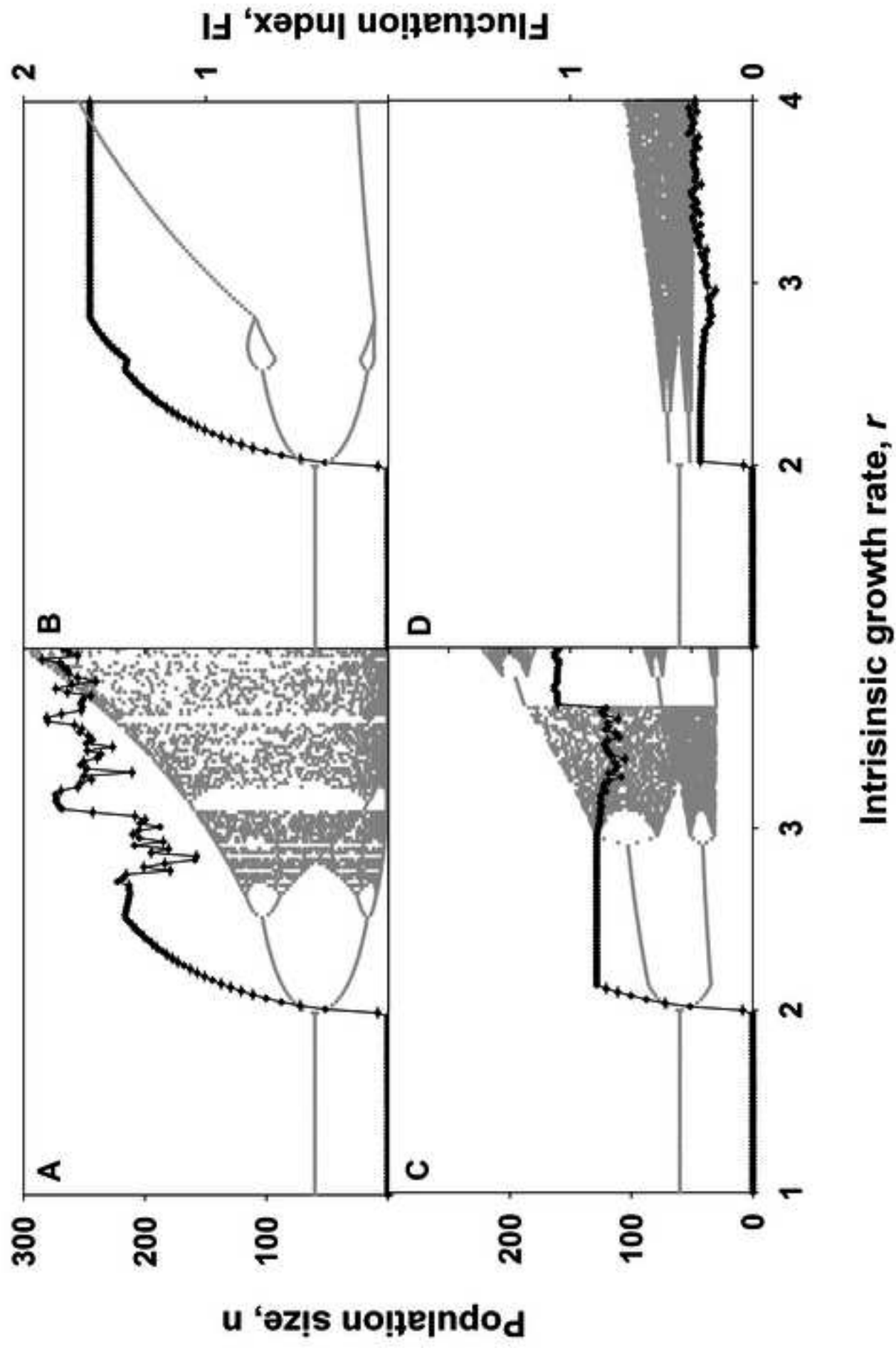

**4. Figure**

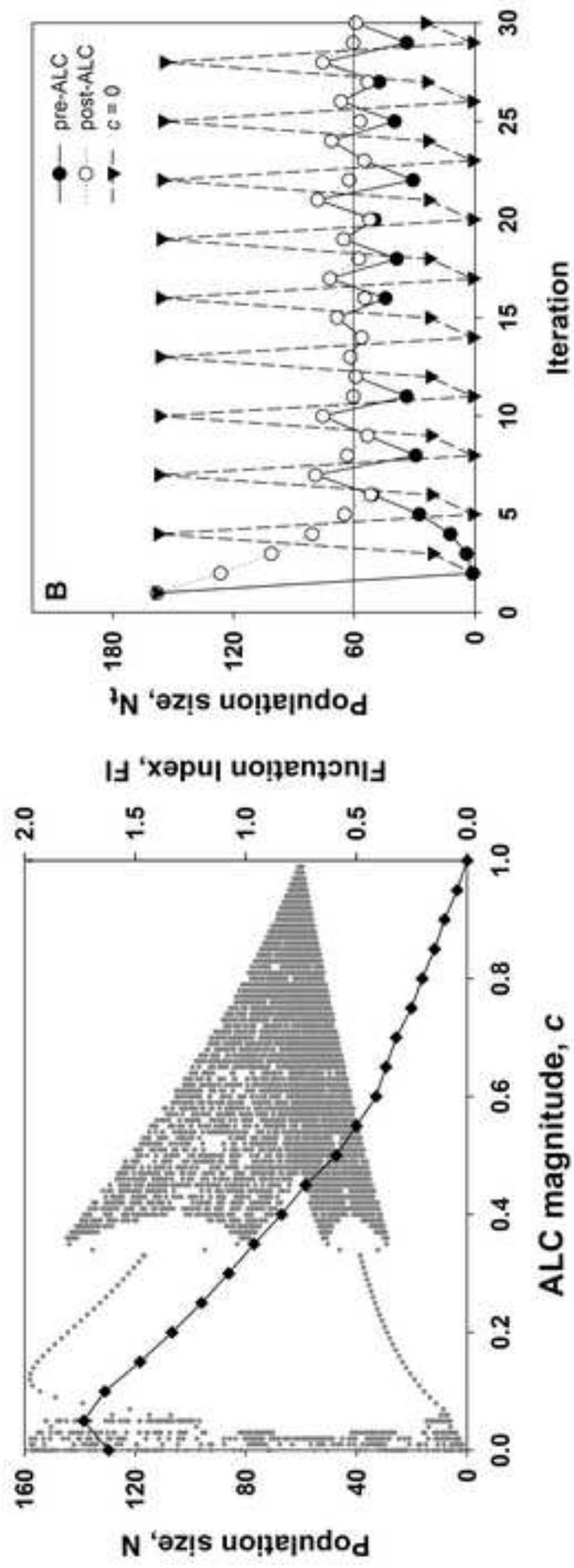

**4. Figure**

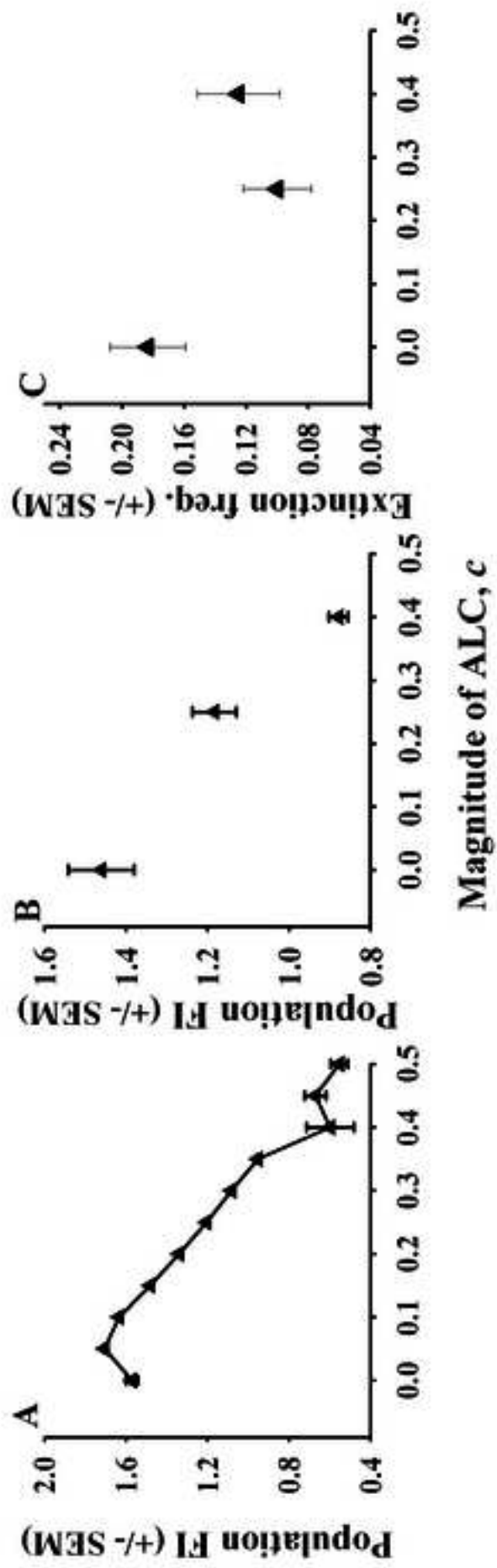

**4. Figure**

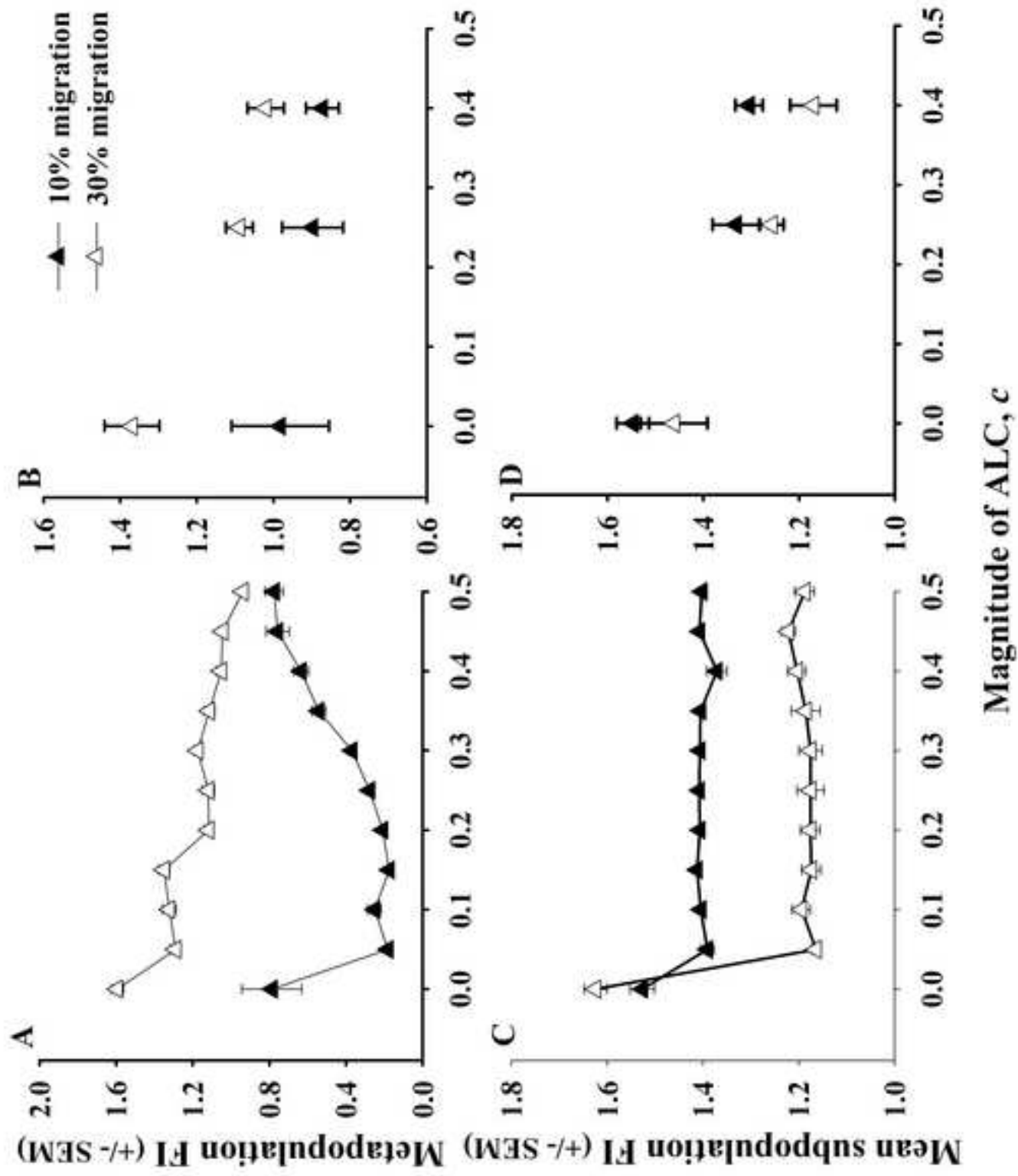

**4. Figure**

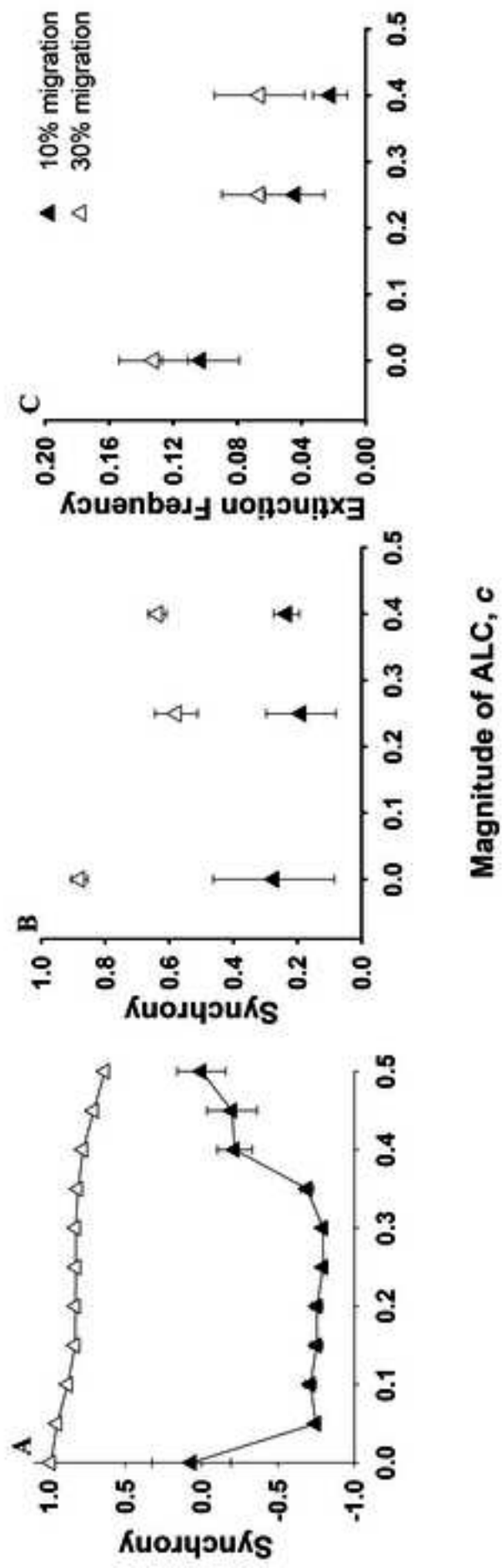

**4. Figure**

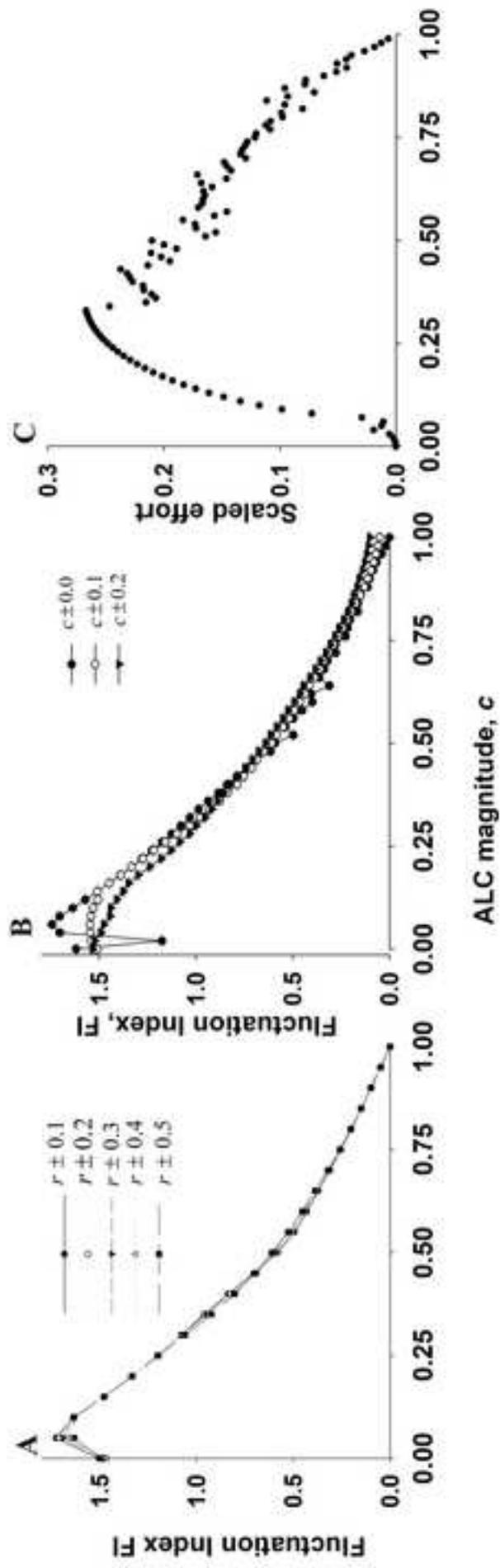